\documentclass{aa}  

\usepackage{graphicx}
\usepackage{float}
\usepackage{txfonts}

\begin{document}

   \title{On The {Orbital Separation Distribution} \newline and Binary Fraction of M Dwarfs}

%   \subtitle{A Log-Normal Fit to the Orbital Separation Distribution}

   \author{Nicholas Susemiehl
          \inst{1}
          \and
         Michael R. Meyer\inst{1}
          }

   \institute{1: Department of Astronomy, University of Michigan,
              Ann Arbor, MI, USA\\
             }

   \date{Accepted September 9, 2021}

% \abstract{}{}{}{}{} 
% 5 {} token are mandatory
 
  \abstract
   {} 
   {We present a new estimate for the binary fraction {(the fraction of stars with a single companion)} for M dwarfs using a log-normal fit to the {orbital separation distribution}.}
   {We use point estimates of the binary fraction {(binary fractions over specific separation and companion mass ratio ranges)} from four M dwarf surveys sampling distinct orbital radii to fit a log-normal function to the {orbital separation distribution}. This model, alongside the companion mass ratio distribution given by \citet{2013A&A...553A.124R}, is used to calculate the frequency of companions over the ranges of mass ratio (q) and orbital separation (a) over which the referenced surveys were collectively sensitive - [0.60 $\leq$ q $\leq$ 1.00] and [0.00 $\leq$ a $\leq$ 10,000 AU]. This method was then extrapolated to calculate a binary fraction which encompasses the broader ranges of [0.10 $\leq$ q $\leq$ 1.00] and [0.00 $\leq$ a < $\infty$ AU]. Finally, the results of these calculations were compared to the binary fractions of other spectral types.}
   {{The binary fraction over the constrained regions of [0.60 $\leq$ q $\leq$ 1.00] and [0.00 $\leq$ a $\leq$ 10,000 AU] was calculated to be $0.229 \pm 0.028$. This quantity was then extrapolated over the broader ranges of q (0.10 - 1.00) and a (0.00 - $\infty$ AU) and found to be $0.462^{+0.057}_{-0.052}$. We used a conversion factor to estimate the multiplicity fraction from the binary fraction and found the multiplicity fraction over the narrow region of [0.60 $\leq$ q $\leq$ 1.00] and [0.00 $\leq$ a $\leq$ 10,000 AU] to be $0.270 \pm 0.111$.  Lastly, we estimate the multiplicity fractions of FGK, and A stars using the same method (taken over [0.60 $\leq$ q $\leq$ 1.00] and [0.00 $\leq$ a $\leq$ 10,000 AU]) and find that the multiplicity fractions of M, FGK, and A stars, when considered over common ranges of q and a, are more similar than generally assumed.}}
   {}


   \maketitle
%
%________________________________________________________________

\section{Introduction}
    Because M dwarfs comprise as much as 75\% of local stellar populations \citep[][and references therein]{2006AJ....132.2360H,2019AJ....157..216W}, it is important to have an accurate census of their properties. Understanding the multiplicity fraction, the fraction of stellar systems with one or more gravitationally-bound companion (regardless of whether it is above or below the hydrogen burning limit), of M dwarfs will have significant implications for star formation theories, inform us about the emergence of planetary systems around low mass stars, and allow for the modeling of both galactic and extra-galactic stellar populations. A related quantity is the binary fraction, the fraction of stars with a single gravitationally-bound companions. {The key difference between these two statistics is that the multiplicity fraction considers systems with more than two companions while the binary fraction does not. Mathematically, these quantities are expressed as $MF = \frac{N_{D}+N_{T}+N_{H}}{N_{S}+N_{D}+N_{T}+N_{H}}$ and $BF = \frac{N_{D}}{N_{S}+N_{D}+N_{T}+N_{H}}$ (where $MF$ denotes the multiplicity fraction, $BF$ the binary fraction, and $N_S$, $N_D$, $N_T$, $N_H$ the number of single, double, triple, and higher-order systems respectively). These differences extend to formation theories:  the multiplicity fraction may be more related to gravitational fragmentation of star forming molecular cloud cores than the more simple binary fraction \citep[cf.][]{2010ApJS..190....1R, 2014MNRAS.437.1216D}. In addition, differences between the multiplicity and binary fractions as a function of primary star mass are another key constraint of formation theory. Multiple conventions exist to define the mass ratio and orbital distribution for higher order systems so we only consider the binary fraction when fitting our model in order to avoid these complications.} Despite its import, the multiplicity properties of M dwarfs are still not yet fully understood. \par
    
    We model the binary fraction using the companion mass ratio distribution ($\psi$) and {orbital separation distribution} ($\phi$). The companion mass ratio distribution represents the fraction of stars which have a companion within a given range of the mass ratio. {The mass ratio, $q$, is defined as the ratio between the mass of a secondary star and the mass of the respective primary star where, by definition, the mass of the primary star is greater than that of the secondary so that $q \leq 1$}. The {orbital separation distribution} is the fraction of stars which have a companion as a function of the separation between the two stars. The semi-major axis of an orbital ellipse, $a$,  serves as a measure of the separation between the primary and secondary stars. Integrating either of these functions over specific ranges of their respective parameters returns the fraction of M dwarfs which have a single companion over that range {if one assumes the other parameter has been integrated over its full parameter space. $\psi$ and $\phi$ must be combined to provide a binary fraction over a fixed ranges of q and a, calculated as}: 
    \begin{equation}
        f = \int_{q_{min}}^{q_{max}}\psi dq * \int_{log_{10}a_{min}}^{log_{10}a_{max}}\phi dlog_{10}(a)
        \label{eq:1}
    \end{equation} 
    where $q_{min}, q_{max}, log_{10}a_{min},$ and $log_{10}a_{max}$ represent the lower and upper bounds for the mass ratio and semi-major {axes} of interest. \par

     The companion mass ratio distribution is discussed in \citet{2013A&A...553A.124R}, who find:
    \begin{equation}    
        \psi = q^{.25 \pm 0.29} 
        \label{eq:2} 
    \end{equation}
    describes this distribution for M dwarfs {as well as field FGK stars and intermediate mass stars in the young ScoOB2 association.} The present work attempts to fit the functional form $\phi$ for the orbital {separation} distribution.\par
    
    This method for calculating the binary fraction is built upon a key assumption: that the companion mass ratio distribution does not depend on orbital separation. This assumption allows for the calculation of binary fraction{s} using Equation (1). \citet{2013A&A...553A.124R} investigate this assumption {for M dwarfs based on the data of \citet{2012ApJ...754...44J}} and are unable to reject it. Furthermore, we only take into consideration binary systems and not triplets, quadruplets, etc. We further discuss these caveats in Section 4. \par
    
    Numerous attempts have been made to constrain the multiplicity of M dwarfs \citep[e.g.][]{1992ApJ...396..178F, 2012ApJ...754...44J, 2019AJ....157..216W}, but none include samples of stellar binary pairs that are complete over the full ranges of mass ratio ({$0 \leq q \leq 1$)} or separation ($0 < a < \infty$). Instead, observational xresults are necessarily limited to specific ranges of parameters.  These past works have made useful constraints to important parameters regarding the population of M dwarfs. For instance, \citet{1992ApJ...396..178F} found that about $42 \pm 9 \%$ of nearby M dwarfs (with mass ratio 0.2<q<1 and separation 0 <a< 10,000 AU) have a companion {and Ward-Duong et al. (2015) estimated a companion star fraction of $23.5\% \pm 3.2\%$ over a similarly wide region of the parameter space (3<a<10,000 AU, 0.2<q<1)}. \citet{2012ApJ...754...44J} suggests that the multiplicity of M dwarfs is intermediate between that of higher-mass stars and brown dwarfs. {\citet{2017A&A...597A..47C} found a peak in the orbital separation distribution of M dwarf multiple system at 2.5 - 7.5 AU, which is similar to the broad peak at around 4-20 AU found by \citet{2019AJ....157..216W} \citep[cf.][]{2021AJ....161...63W}}.\par
    
    In Section 2 of this paper, we describe the methods used to fit the {orbital separation distribution} and calculate the binary fraction. In Section 3, the results are presented. In Section 4, we discuss the implications and further investigate the assumptions used. Section 5 summarizes our work.\par

%__________________________________________________________________

\section{Methods}
\subsection{The data}
    The first step in exploring the orbital {separation} distribution and binary fraction of M dwarfs was to synthesize point estimates of the binary fraction from a variety of different M dwarf surveys. Data was compiled from surveys which employed the radial velocity \citep{1999A&A...344..897D} and direct imaging \citep{2017A&A...597A..47C, 2012ApJ...754...44J, 2015MNRAS.449.2618W} companion detection methods. This allowed us to investigate companions over a broad range of semi-major axes. The projected separations from the direct imaging surveys {of} \citet{2017A&A...597A..47C} and \citet{2015MNRAS.449.2618W} were converted to physical separations using the correction factor described in \citet{1992ApJ...396..178F} of 1.26 (Janson et. al. 2012 already made a correction {using this factor}). {Other works, including \citet{2011ApJ...733..122D}, suggest alternative forms to this correction factor that account for other orbital parameters such as eccentricity, but the correction factor estimated by \citet{2011ApJ...733..122D} ($1.16^{+0.81}_{-0.31}$) is similar to the 1.26 suggested by \citet{1992ApJ...396..178F}.}\par
    
   We include only detected multiple systems associated with values of mass ratio and semi-major axis over which each survey was at least 90\% complete. The lower limit on mass ratio considered was set by \citet{2017A&A...597A..47C} as q $\geq$ 0.60. Any detected multiple system which was separated by a distance outside the range that each particular survey was at least 90\% complete was also removed {(i.e. no longer considered a detection of a binary system}). We also excluded any higher order systems by dropping such systems from what we consider to be companions but including them in the total {primary} star number, {effectively treating these systems as single stars}. \citet{1999A&A...344..897D} detected six higher order systems, \citet{2017A&A...597A..47C} detected four, \citet{2012ApJ...754...44J} detected 14, and \citet{2015MNRAS.449.2618W} detected five.  The binary fraction point estimates for each survey were calculated by dividing the remaining number of binary pairs by the total parent population of each survey. {The error bars on the binary fraction point estimates were taken to be the Poisson counting error of the respective surveys (calculated after the aforementioned cleaning steps)}. Table (1) gives the point estimate of the binary fraction for each survey along with the respective detection method and range of semi-major axis. {Similarly, Table (2) depicts the total number of parent stars surveyed and the number of companion detections for each reference.}\par

  In certain cases, assumptions had to be made about details of the surveys. \citet{1999A&A...344..897D} reported the period and orbital velocity of stars in their sample, but not the semi-major axis. Kepler's Third Law and the masses of the primaries were used to convert the given values into semi-major axes. \citet{2012ApJ...754...44J} did not note the minimum mass ratio that the survey could detect. The authors state: "In every case where $q_m * m_a < 0.08$ $M_{sun}$, the system is removed from the analysis... Furthermore, we include only stars with primary mass > 0.2 $M_{sun}$, since lower mass stars are not fully complete out to 52 pc" \citep{2012ApJ...754...44J}. From this, we took a minimum secondary mass of 0.08 $M_{sun}$ and a minimum primary mass of 0.2 $M_{sun}$ {to calculate a minimum mass ratio of 0.4.} Both \citet{2012ApJ...754...44J} and \citet{2015MNRAS.449.2618W} included a small number of multiple systems with K-type primary stars, each of which was removed from this analysis. The data from \cite{2015MNRAS.449.2618W} was split into two bins by orbital separation: 0 - 100 AU and 100 - 10,000 AU because it covers such a wide range of semi-major axes (3 - 10,000 AU). \par
  
 In order to justify our use of the \citet{2013A&A...553A.124R} power law (Equation 2), we employed the Kolmogorov–Smirnov (KS) test to compare the mass ratio samples of Ward-Duong et al. (2015) and Cortes-Contreras et al. (2016) to ones drawn from the mass ratio power law distribution of \citet{2013A&A...553A.124R} (considering only q>0.6; Fischer \& Marcy 1992 and Janson et al. 2012 were used to fit the power law of Reggiani and Meyer 2013 and so were not tested here). {These KS tests returned p-values of 0.04 and 0.02 respectively. Based on these p-values, we do not {reject} the null hypothesis that there is no difference between the distributions of Ward-Duong et al. (2015) and Cortes-Contreras et al. (2016) and the mass ratio distribution proposed by \citet{2013A&A...553A.124R} at the 0.01 significance level}. As an alternative, future similar work could consider using the mass ratio distribution offered by \citet{2019MNRAS.489.5822E} {which suggests a broken power law fit and found an excess of near-equal mass systems. Using this mass ratio distribution would likely change the results found in this study.} \par
        \begin{table*}[t]
        
        \caption{Binary Fraction Point Estimates, q $\geq$ 0.6}
        \label{table:1}
        \centering
        \begin{tabular}{c c c c }
        \hline\hline
        Reference & Semi-Major Axis Range (AU) & Primary Mass Range ($M_{sun}$) & Binary Fraction Estimate  \\    
        \hline 
    	Delfosse et al. (1999)                        & 0.00 - 4.63           & 0.13 - 0.44                 & 0.04 $\pm$ 0.018 \\
        Cortés-Contreras et al. (2017)                & 3.28 - 37.17*           & 0.14 - 0.49                 & 0.07 $\pm$ 0.012 \\
        Janson et al. (2012)                          & 3.00 - 227.00            & 0.20 - 0.59                 & 0.17 $\pm$ 0.015 \\
        Ward-Duong et al. (2015) A                    & 3.78 - 126.00*            & 0.24 - 0.67                 & 0.11 $\pm$ 0.022 \\
        Ward-Duong et al. (2015) B                    & 126.00 - 12,600.00*         & 0.32 - 0.52                 & 0.07 $\pm$ 0.017 \\
        \hline 
        \end{tabular}
        \tablefoot{This table depicts each of the M dwarf surveys used in this study alongside their respective ranges of semi-major axis over which the survey was at least 90\% complete, ranges of primary star masses, and the point estimates of the binary fraction that we made after excluding companion detections outside the respective semi-major axis ranges and with q < 0.6. \newline * Projected separations converted into physical separations by multiplying limits by 1.26 \citep{1992ApJ...396..178F}.}
    \end{table*}

 \begin{table*}[t]
        
        \caption{{Survey Details}}
        \label{table:2}
        \centering
        \begin{tabular}{c c c c}
        \hline\hline
        Reference & Total Num. of Stars Surveyed & Total Num. of Detected Companions & {Companion Subsample} \\    
        \hline 
    	Delfosse et al. (1999)                        & 127 & 12          & 5                  \\
        Cortés-Contreras et al. (2017)*                & 490 & 80           & 34               \\
        Janson et al. (2012)                          & 701 & 219            & 117          \\
        Ward-Duong et al. (2015) A                    & 232 & 65            & 25 \\
        Ward-Duong et al. (2015) B                    & 232 & 65     & 16 \\
        \hline 
        \end{tabular}
        \tablefoot{{This table summarizes each of the M dwarf surveys used in this study. The first column lists the total number of type M primary stars included in each survey. The second column is the number of companions detected by each survey. The third column is the number of companions remaining after the corrections described in Section 2.1 are made. The binary fraction estimates described in Table 1 can be found by dividing the fourth column of this table by the second.\newline *:We used the volume-limited sub-sample of Cortés-Contreras et. al. (2017).}}
    \end{table*}

    \subsection{Fitting the model}
    We fit a log-normal model to the orbital separation distribution of the M dwarf binary systems by using the binary fraction point estimates from the referenced surveys (Table 1) as benchmarks to those computed using Equation (1). This model {(Equation 3)} has as free parameters an amplitude (A) which normalizes the function, a mean ($\mu$) which indicates the peak of the distribution, and a standard deviation ($\sigma$) which describes the spread of the distribution. 
    \begin{equation}
       \phi = A * \frac{e^{-(log_{10}(a) - \mu)^{2} / (2\sigma^{2})}}{\sigma*\sqrt{2\pi}} \label{eq:3}
    \end{equation} 
     In order to fit the model and find the best values for these parameters, a Markov Chain Monte Carlo (MCMC) routine was utilized {via the emcee Python package \citep{2013PASP..125..306F}}. Model binary fractions, {which are calculated via Equation (1) by inputting parameter values into $\phi$,}  were compared to the { binary fraction point estimates from the referenced surveys} using a chi-squared log-likelihood function. Uniform log-priors were chosen with bound $0.0 \leq A \leq 2.0$, $-1.0 \leq \mu \leq 2.0$, and $-1.0 \leq \sigma \leq 2.0$ (the latter two given in the log of the semi-major axis in AU). The MCMC was initialized with 100 walkers positioned randomly within the bound of the log-prior and run for 5,000 steps. After this, the flattened chains of each of the three parameters were plotted in order to visualize the convergence of the walkers. This informed the decision to discard the first 200 steps of each flatchain as the burn-in values. The best fit amplitude, mean, and standard deviation were then calculated as the median value of these chains, with error corresponding to the 68\% confidence interval of the respective posterior distributions. A cornerplot (Figure 1) was produced which shows the distributions of each of the three parameters. 
    \begin{figure}[H]
        \includegraphics[width=0.5\textwidth]{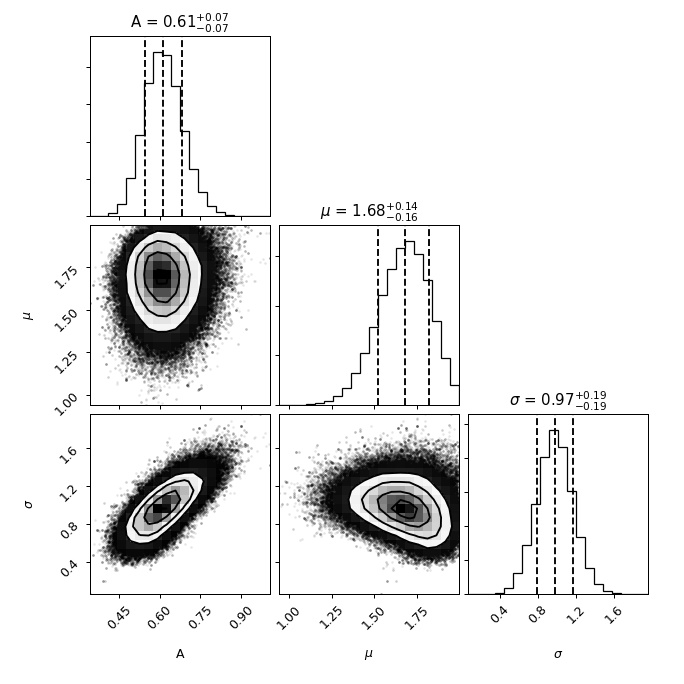}
        \caption{Marginal distributions of each of the three parameters as well as the covariances between all of them. The median values and 68\% confidence intervals of each of the parameters are drawn as vertical lines {and also noted above the respective histograms}.}
    \end{figure}
    \subsection{Calculating the binary fraction}
    Next, the model for the orbital {separation} distribution, with its best-fitting parameters, was used alongside the companion mass ratio distribution model from \citet{2013A&A...553A.124R} to calculate the binary fraction of M dwarfs as: 
    \begin{equation}
    \begin{split}
        f = \int_{q_{min}}^{q_{max}}q^{.25} dq * \\ A_{best} *  \int_{log_{10}a_{min}}^{log_{10}a_{max}} \frac{e^{-(log_{10}(a) - \mu_{best})^{2} / (2\sigma_{best}^{2} )}}{\sigma_{best}*\sqrt{2\pi}} dlog_{10}(a)
    \end{split}
    \label{eq:4}
    \end{equation} 
    This calculation resulted in a binary fraction that is representative over these limited ranges of mass ratio and semi-major axis. These ranges were later expanded to [0.10 $\leq$ q $\leq$ 1.0] and [0 $\leq$ a < $\infty$ AU] to allow for the calculation of a broader M dwarf binary fraction. The error on the binary fraction was calculated as the 68\% confidence interval of the probability distribution function of the binary fraction, which was generated by sampling 1,000 values of A, $\mu$, and $\sigma$ from the posterior space and calculating a value of the binary fraction for each of these parameter values (as per Equation 4). Figure (2)  summarizes the model's fit to the data.
    \begin{figure}[H]
        \includegraphics[width=0.5\textwidth]{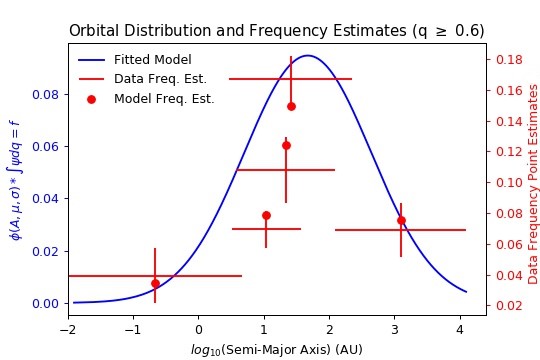}
        \caption{Comparison of the binary fraction survey data to the fitted model. Each of the red horizontal lines represent one of the five binary fraction estimates from the literature. The right vertical axis position of these lines is the binary fraction reported in the data and the span along the horizontal axis of these lines is the range in semi-major axes that the respective data cover. The red vertical lines are the error in these binary fraction estimates. The blue curve is the fitted log-normal model of the {orbital separation distribution}. Each red dot is the binary fraction estimate obtained by integrating the fitted model over the range of semi-major axis respective to each referenced survey. {When visually assessing the validity of our model, the most important aspect of this plot to notice is that each red data point fits within each red vertical line, implying that our model replicates the referenced binary fraction point estimates within error. }}
    \end{figure}
%______________________________________________________________
 \section{Results}
    \subsection{Orbital separation model}
    The best-fit to point estimates of the binary fraction from the four M dwarf surveys resulted in the parameters of A, $log_{10}(\mu)$, and $log_{10}(\sigma)$, being $0.61 \pm 0.07$, $1.68^{+0.14}_{-0.16}$, and $0.97 \pm 0.19$ respectively. This fit is associated with a reduced chi-squared parameter value of 1.315. With two degrees of freedom, the probability of achieving a value greater than or equal to 1.315 is 0.269. Therefore, we could not reject the null hypothesis that the data came from this model.
    \subsection{Binary fraction}
    Integrating Equation (4) over the constrained regions of [0.60 $\leq$ q $\leq$ 1.00] and [0 $\leq$ a $\leq$ 10,000 AU] resulted in a binary fraction of $0.229 \pm 0.028$. {Equation (4) was then integrated again over [0.1 $\leq$ q $\leq$ 1.0] and [0.0 $\leq$ a < $\infty$ AU] to calculate a broad binary fraction of $0.462^{+0.057}_{-0.052}$}. This binary fraction considers stellar and sub-stellar companions to M dwarfs, but not planetary companions. {The binary fraction corresponding to mass ratios and semi-major axes outside of the ranges of [0.60 $\leq$ q $\leq$ 1.00] and [0 $\leq$ a $\leq$ 10,000 AU] must be considered with care because they are extrapolations of our model.}
    {
    \subsection{Multiplicity fraction}
    Due to its import in constraining theories of stellar formation \citep[][and others]{2012ApJ...754...44J, 2014MNRAS.437.1216D, 2019AJ....157..216W}, we provide an estimate of the multiplicity fraction based on our binary fraction fit. \citet{2019AJ....157..216W} estimates the ratio of multiple systems (including binary and higher order systems) to binaries to be 1.179 $\pm$ 0.107 (with error propagated from the Poisson error in the number of multiple and binary systems). By assuming that the binary fraction and the total number of higher order systems are independent, we were able to multiply our binary fraction by this ratio to obtain a multiplicity fraction estimate of $0.270 \pm 0.111$ over the constrained regions of [0.60 $\leq$ q $\leq$ 1.00] and [0 $\leq$ a $\leq$ 10,000 AU]. The multiplicity of M dwarfs integrated over the same region of mass ratio but over all values of separation (0-$\infty$) is presented here for use in the comparison made in section 4.2 to be 0.272 $\pm$ 0.111. 
    }

\section{Discussion}
\subsection{Comparisons to other M dwarf surveys}
We compare our model to the to the {observation} of {\citet{2015ApJS..216....7B}}, which estimated the frequency of brown dwarf companions to M dwarf hosts. They {observe} the frequency of brown dwarf companions to M dwarf primaries over a semi-major axis range of 10 - 100 AU and mass ratio range of 0.039 - 0.224 to be $0.028^{+0.024}_{-0.015}$. We integrated Equation (1) with the best fit parameters described above over these ranges of semi-major axis and mass ratio  and obtained an binary fraction estimate of $0.027^{+0.010}_{-0.007}$ {\citep[cf.][]{2012AJ....144...64D}}. While the sample of \citet{2015ApJS..216....7B} is drawn from nearby young moving groups and our preceding work was built using samples drawn from field stars, this comparison is justified because the multiplicity of stars in these moving groups does not appear to be discernibly different from the field \citep[see][]{2015A&A...580A..88E, 2019ApJ...886...95D}. {None of the surveys used to construct our binary fraction model probed sub-stellar companions, and so the capability of our model to reproduce the results of \citet{2015ApJS..216....7B} is of particular note.}
\par \citet{2019AJ....157..216W} ({W19}) also studied the {orbital separation distribution} and binary fraction of M dwarfs. They find {an upper limit to the} peak in the this distribution {to be at} 20 AU {(projected separation)} and note that most stellar companions to M dwarfs lay between 4 and 20 AU. {When corrected to a physical separation using the 1.26 correction factor of Fischer \& Marcy (1992) (Section 2.1), W19's peak to the orbital separation distribution becomes 25 AU. This falls about 1.9 standard deviations from our peak estimate of $48^{+18}_{-15}$ AU (our orbital separation model estimates the probability of the peak of this distribution occurring at less than 20 AU to be about 3\%.}) An attempt was made to draw a comparison between our binary fraction and {that} of {W19}. {However, it was not straight-forward to find the range of q and separation over which the survey was complete. We therefore attempted to estimate this ourselves.} First, we calculated a binary fraction from {the W19} sample by using their stated number of single:double:triple:high-order systems. We assumed that the {binary} fraction {uniformly} covered mass ratios from a minimum q set by the hydrogen burning limit divided by the median primary mass of their sample to a maximum q of 1. We assumed the companion mass ratio distribution of \citet{2013A&A...553A.124R} to convert this to the range of mass ratio that is representative of our binary fraction, 0.6 - 1.0. Furthermore, we assumed that the multiplicity fraction of {W19} was representative for companions with separations less than 10,000 AU. However, the binary fraction that we estimated from the sample of {W19}, {0.133 $\pm$ 0.014} did not fall within error of our constrained binary fraction, {0.229 $\pm$ 0.028}. This discrepancy is likely due to the numerous assumptions we were forced to make.
\par {Another survey of M dwarf multiple systems, \citet{2015ApJ...813...75S}, finds the fraction of M dwarf companions with periods less than 90 days to M dwarf hosts to be $0.11^{+0.02}_{-0.04}$. This estimate is much higher than Delfosse et al. for closely-separated M dwarfs and would have a significant impact on our fit if it were included as a data point. However, we were unable to include \citet{2015ApJ...813...75S} in our analysis because the orbital separation and mass ratio completeness limits and values for each system are not easy to ascertain. \citet{2013MNRAS.429..859J} was also considered for inclusion in our fit but it was also not clear over what range of mass ratios and orbital separations to which the survey was complete.} {The microlensing survey \citet{2016MNRAS.457.4089S} was not used in this analysis because there were difficulties in translating this work into usable point estimates without making strong assumptions that were not able to verify regarding their sensitivity limits in mass ratio and orbital separation. The inclusion of microlensing surveys would improve our analysis significantly because of their sensitivity to low-separation companions}
\subsection{Comparisons to other spectral type binary fractions}
We now compare the binary fraction of M dwarfs to that of sun-like FGK- and more massive A-type stars over the range of mass ratios studied here but over all separations (0.6 $\leq$ q $\leq$ 1.0 and 0.00 $\leq$ a $\leq$ $\infty$ AU). We expand this integration to be over all separations because of the greater prevalence of higher mass stars at larger separations. In order make this comparison, we choose an appropriate form of the companion mass ratio distribution for the stellar hosts in question then recalculate the orbital separation distribution for each population of stars based on previous works. To compare to the FGK binary fraction, we extrapolated the model of orbital separation distribution from \citet{2010ApJS..190....1R} {(which peaks at 48 AU)} to integrate it over separations from 0 to $\infty$ AU. This was used alongside the companion mass ratio distribution of Reggiani and Meyer (2013) integrated from 0.6 to 1.0 to find an FGK-star binary fraction of 0.17 $\pm$ 0.03, where the error is estimated as the Poisson counting error based on the survey of \citet{2010ApJS..190....1R}. Following the process in Section 3.3 of this work, we convert this binary fraction to a multiplicity fraction using the ratio of multiple systems to binary systems described in \citet{2010ApJS..190....1R} (1.33 $\pm$ 0.14) and find an FGK-star multiplicity fraction of $0.23 \pm 0.14$ over 0.6 $\leq$ q $\leq$ 1.0 and 0.00 $\leq$ a $\leq$ $\infty$ AU. We made a similar calculation for the A star binary fraction, this time based on \citet{2014MNRAS.437.1216D} {the orbital separation distribution of this survey has a mean at nearly 400 AU)}. However, because the A star sample of \citet{2013A&A...553A.124R} (young stars in an association) is  different than that of \citet{2014MNRAS.437.1216D} (field stars) we chose to instead use the companion mass ratio distribution distribution described in \citet{2014MNRAS.437.1216D} to make this calculation (which is similar to that of \citet{2013A&A...553A.124R} at close separations but not at separations greater than 125 AU). Using this, we found an A star binary fraction of $0.11 \pm 0.02$ and multiplicity fraction of $0.15^{+0.25}_{-0.15}$ (by multiplying the binary fraction by the ratio of multiple systems to binary systems of \citet{2014MNRAS.437.1216D}, 1.36 $\pm$ 0.25) over mass ratio 0.6 $\leq$ q $\leq$ 1.0 and orbital separation 0.00 $\leq$ a $\leq$ $\infty$ AU. The binary fractions calculated here decrease as spectral type increases, but the multiplicity fractions remain consistent (with large error bars). {These calculations are built on a number of assumptions. Most notably, that the orbital separation and companion mass ratio distributions used are complete over the entire space of 0.6 $\leq$ q $\leq$ 1.0 and 0.00 $\leq$ a $\leq$ $\infty$ AU (the orbital separation distribution of \citet{2014MNRAS.437.1216D} is incomplete at low separations). Also that the multiplicity fraction can be estimated from the binary fraction as we have done above. Importantly, these results must be considered with these assumptions and with their large error bars in mind.} This finding is in contrast to \citet{2013ARA&A..51..269D}, which suggests that the occurrence rate of companions increases with spectral type. Nonetheless, when compared over common ranges of mass ratios, the multiplicity fraction as a function of spectral type is more similar than commonly appreciated.\par
\subsection{Additional caveats}
The binary fraction study conducted here was only concerned with the presence of binary systems within the M dwarf population. No considerations were made to account for the presence of higher order (triple, quadruple, etc.) systems {during the fitting process. Instead, we converted the binary fraction to a multiplicity fraction using the empirical findings of other works}. While such systems do exist, they are relatively rare. A recent M dwarf multiplicity study found that only 3.3\% of M dwarfs exist as triple and higher order systems \citep{2019AJ....157..216W}. {However, it appears that higher order systems are more common around higher mass stars \citep{2010ApJS..190....1R, 2015MNRAS.449.2618W}.}\par

Our work assumes that the mass ratio of a stellar binary system does not depend on the separation between its components \citep{2013A&A...553A.124R}. \citet{2017ApJS..230...15M} found that the distributions of mass ratio and period (which is directly proportional to separation) of O- and B-type main sequence primaries are not independent. {If this is also true for M dwarf binaries, then Equation (1) would not hold. For the time being, we accept the findings of \citet{2013A&A...553A.124R} but note that} further work exploring the interdependence of mass ratio and orbital separation for specifically type M stars {is needed}.

Originally, our fit to the orbital separation distribution utilized the multiplicity estimate from {the radial velocity data of \citet{1992ApJ...396..178F}}. The inclusion of this data point (binary fraction of 0.083 $\pm$ 0.03 over 0.6 $\leq$ q $\leq$ 1 and 0.04 $\leq$ a $\leq$ 4) {gave a worse fit and} led to a higher reduced chi-squared of 2.143. We believe  {this is because} the primary mass range of the sample of \citet{1992ApJ...396..178F}, 0.14 - 0.33 $M_{Sun}$, does not include higher mass M dwarfs present in the other referenced surveys. {While this primary mass range may not seem substantially different than that of Delfosse et al. (1999), we examined the spectral type distributions of the parent populations of \citet{1992ApJ...396..178F} and Delfosse et al. (1999), \citep[][respectively]{1989ApJ...344..441M, 1998A&A...331..581D}, and found that they are distributed significantly differently (i.e. Marcy \& Benitz 1989, and therefore Fischer \& Marcy 1992, include more low mass stars than Delfosse et al. 1998/Delfosse et al. 1999) by performing a KS test and receiving a p-value of 0.0007}. {Previous work suggests the companion mass ratio distribution for brown dwarfs is less flat than that of M dwarfs \citep{2007prpl.conf..427B, 2018MNRAS.479.2702F}. The inclusion of \citet{1992ApJ...396..178F} may have been problematic because its sample is closer to the brown dwarf regime than the other referenced surveys.}  Therefore, we chose to exclude this point from our model fit. \par

\section{Summary}
We find that the distribution of M dwarf binary separations peaks between {33-66} AU within a 68 \% confidence interval (23-85 AU with 95 \% confidence), which is {comparable to the peak for FGK stars but }smaller than the peak for A stars \citep{2010ApJS..190....1R, 2014MNRAS.437.1216D}. {Using the model built from this finding as well as the companion mass ratio distribution given by Reggiani \& Meyer (2013)}, we find that the binary fraction over [0.60 $\leq$ q $\leq$ 1.00] and [0.00 $\leq$ a $\leq$ 10,000 AU] for M dwarfs to be  $0.229 \pm 0.028$ {(this is the most robust of our findings as it required the fewest extrapolations)}. {We used this to derive} the multiplicity fractions over the same range of mass ratio and over all separations for M, FGK, and A stars of $0.27 \pm 0.11$, $0.23 \pm 0.14$, and $0.15 \pm 0.25$ respectively. {We note that these three values suggest that the the multiplicity fraction is similar over these three spectral types and ({within these ranges of mass ratio and orbital separation}) but the error bars on these values are large}. Furthermore, we find the binary fraction of M dwarfs over 0.10 $\leq$ q $\leq$ 1.00 and 0.00 $\leq$ a $\leq$ $\infty$ AU to be $0.462^{+0.057}_{-0.052}$, {implying that nearly half of all M dwarfs {(over all q and a)} have a stellar or sub-stellar companion. This assertion is highly dependent on the validity of the assumptions we made. Namely, that the frequency of binary companions is consistent beyond the parameter space of the surveys used to build our model.} {Data which would help improve the predictive power of our orbital separation model may include M dwarf multiplicity surveys which probe pairs with separations between 3 and 30 AU {and low mass ratios}. }Future studies may expand upon this work by surveying the binary fraction of M dwarf companions in the brown dwarf regime, explore correlations between the companion mass ratio and orbital separation distributions, as well as compare M dwarf binary fraction to L, T, and Y- Dwarf multiplicities. 

\begin{acknowledgements}
      We thank the Formation and Evolution of
Planetary Systems group for their consistent
support, Dr. Max Moe and Dr. Kimberly Ward-Duong for their insightful advice, and an anonymous referee whose helpful reviews significantly improved the manuscript.
\end{acknowledgements}

%-------------------------------------------------------------------

\bibliographystyle{aa.bst} % style aa.bst
\bibliography{ref.bib} % your references Yourfile.bib

\end{document}